\documentclass[conference]{IEEEtran}

\IEEEoverridecommandlockouts
\usepackage{cite}
\usepackage{amsmath,amssymb,amsfonts}
\usepackage{algorithmic}
\usepackage{graphicx}
\usepackage{amssymb}
\usepackage{subcaption}
\usepackage{pifont}
\usepackage{multirow}
\usepackage{booktabs}
\usepackage{makecell}
\usepackage{textcomp}
\usepackage{enumerate}
\usepackage{xcolor}
\usepackage{xspace}
\usepackage{wrapfig,lipsum,booktabs}

\def\BibTeX{{\rm B\kern-.05em{\sc i\kern-.025em b}\kern-.08em
    T\kern-.1667em\lower.7ex\hbox{E}\kern-.125emX}}

\newcommand{\Name}{\textsc{Mercury}\xspace}

\begin{document}

\title{\Name: An Automated Remote Side-channel Attack to Nvidia Deep Learning Accelerator\\
}

\author{
    \IEEEauthorblockN{Xiaobei Yan\IEEEauthorrefmark{1}, Xiaoxuan Lou\IEEEauthorrefmark{1}, Guowen Xu\IEEEauthorrefmark{1},Han Qiu\IEEEauthorrefmark{3},Shangwei Guo\IEEEauthorrefmark{4},Chip Hong Chang\IEEEauthorrefmark{2},Tianwei Zhang\IEEEauthorrefmark{1}}
    \IEEEauthorblockA{\IEEEauthorrefmark{1} School of Computer Science and Engineering, Nanyang Technological University, Singapore}
    \IEEEauthorblockA{\IEEEauthorrefmark{2} School of Electrical and Electronic Engineering, Nanyang Technological University, Singapore}
        \IEEEauthorblockA{\IEEEauthorrefmark{3} Institute for Network Sciences and Cyberspace, Tsinghua University, China}
            \IEEEauthorblockA{\IEEEauthorrefmark{4} College of Computer Science, Chongqing University, China}
    \IEEEauthorblockA{\{xiaobei002,xiaoxuan001\}@e.ntu.edu.sg, guowen.xu@ntu.edu.sg,\\ qiuhan@tsinghua.edu.cn,swguo@cqu.edu.cn,ECHChang@ntu.edu.sg,tianwei.zhang@ntu.edu.sg}
}
\maketitle
\begin{abstract}
   DNN accelerators have been widely deployed in many scenarios to speed up the inference process and reduce the energy consumption. One big concern about the usage of the accelerators is the confidentiality of the deployed models: model inference execution on the accelerators could leak side-channel information, which enables an adversary to preciously recover the model details. Such model extraction attacks can not only compromise the intellectual property of DNN models, but also facilitate some adversarial attacks. 

    Although previous works have demonstrated a number of side-channel techniques to extract models from DNN accelerators, they are not practical for two reasons. (1) They only target simplified accelerator implementations, which have limited practicality in the real world. (2) They require heavy human analysis and domain knowledge. To overcome these limitations, this paper presents \Name, the first automated remote side-channel attack against the off-the-shelf Nvidia DNN accelerator. The key insight of \Name is to model the side-channel extraction process as a sequence-to-sequence problem. The adversary can leverage a time-to-digital converter (TDC) to remotely collect the power trace of the target model's inference. Then he uses a learning model to automatically recover the architecture details of the victim model from the power trace without any prior knowledge. The adversary can further use the attention mechanism to localize the leakage points that contribute most to the attack. Evaluation results indicate that \Name can keep the error rate of model extraction below 1\%.
\end{abstract}

\section{Introduction}
\label{sec:intro}
Modern deep learning technology exhibits a computationally intensive trend to perform more complex tasks. This leads to the popularity of adopting specific hardware to accelerate computation and reduce energy consumption. Field Programmable Gate Arrays (FPGAs) are a prevalent choice for implementing DNN accelerators, and have been widely deployed in large-scale datacenters by various cloud providers, such as Amazon EC2 F1 \cite{amazon} and Microsoft Catapult \cite{putnam2014reconfigurable}. 

However, DNN accelerators in the cloud pose new security challenges. Generally, cloud providers adopt the multi-tenancy policy that facilitates multiple users to share the same FPGA board to enhance the resource utilization \cite{mbongue2020architecture}. Although the circuits of different users can be logically separated, they still share the same power distribution network (PDN). A prior study \cite{gnad2016analysis} shows that the supply voltage at different locations of a PDN is not constant and depends on
the activity of the logic. Therefore, the voltage fluctuation becomes a critical side channel to leak sensitive information across different parts of the FPGA. 
The adversary can abuse the multi-tenancy feature to \textit{remotely} launch an on-chip monitor on the same board with the victim's deep learning model and steal its information \textit{without requiring physical access to the hardware}. It has been commonly referred to as remote side-channel attacks in prior works \cite{tian2021remote,moini2021remote,gravellier2021remote}, exhibiting more flexibility and practicality than physical attacks \cite{yoshida2019model,li2021power,yu2020deepem,hua2018reverse,gupta2023ai}. 


\begin{table}[t]
  \centering
  \caption{Summary of existing works}
  \vspace{-5pt}
  \resizebox{\linewidth}{!}{
    \begin{tabular}{llllllll}
    \Xhline{1.5pt}
    \multicolumn{1}{l|}{\textbf{Attack}} & \multicolumn{1}{l|}{\textbf{Impl.}} & \multicolumn{1}{l|}{\textbf{Model}}& \multicolumn{1}{l|}{\textbf{Target}} & \multicolumn{1}{l|}{\textbf{Aim}} &\multicolumn{1}{l|}{\textbf{Remote}} & \multicolumn{1}{l|}{\textbf{Automated}} & \multicolumn{1}{l}{\textbf{\# of runs}} \\
    \Xhline{1.5pt}
    \multicolumn{1}{l|}{\cite{tian2021remote}} & \multicolumn{1}{l|}{VTA\cite{moreau2019hardware}}& \multicolumn{1}{l|}{CNN} & \multicolumn{1}{l|}{Layer}& \multicolumn{1}{l|}{A} & \multicolumn{1}{l|}{\checkmark} & \multicolumn{1}{l|}{\ding{53}} & \multicolumn{1}{l}{50} \\
    \multicolumn{1}{l|}{\cite{yoshida2019model}}  & \multicolumn{1}{l|}{Home-grown }& \multicolumn{1}{l|}{MLP}& \multicolumn{1}{l|}{Model}& \multicolumn{1}{l|}{W} & \multicolumn{1}{l|}{\ding{53}} & \multicolumn{1}{l|}{\ding{53}} & \multicolumn{1}{l}{60K} \\

    \multicolumn{1}{l|}{\cite{li2021power}} & \multicolumn{1}{l|}{Home-grown} & \multicolumn{1}{l|}{CNN}& \multicolumn{1}{l|}{Multiplication}& \multicolumn{1}{l|}{W} & \multicolumn{1}{l|}{\ding{53}} & \multicolumn{1}{l|}{\ding{53}} & \multicolumn{1}{l}{40K} \\
    \multicolumn{1}{l|}{\cite{yu2020deepem}}  & \multicolumn{1}{l|}{Home-grown}& \multicolumn{1}{l|}{BNN}& \multicolumn{1}{l|}{Model}& \multicolumn{1}{l|}{A,W} & \multicolumn{1}{l|}{\ding{53}} & \multicolumn{1}{l|}{A:\ding{53}; W:\checkmark} & \multicolumn{1}{l}{10K} \\
    \multicolumn{1}{l|}{\cite{hua2018reverse}} & \multicolumn{1}{l|}{Home-grown }& \multicolumn{1}{l|}{CNN} & \multicolumn{1}{l|}{Model}& \multicolumn{1}{l|}{A,W} & \multicolumn{1}{l|}{\ding{53}} & \multicolumn{1}{l|}{\ding{53}} & \multicolumn{1}{l}{--} \\
    \multicolumn{1}{l|}{\cite{gupta2023ai}}  & \multicolumn{1}{l|}{NVDLA}& \multicolumn{1}{l|}{CNN}& \multicolumn{1}{l|}{Layer}& \multicolumn{1}{l|}{A} & \multicolumn{1}{l|}{\ding{53}} & \multicolumn{1}{l|}{\ding{53}} & \multicolumn{1}{l}{--} \\
    \multicolumn{1}{l|}{\cite{zhang2021stealing}} & \multicolumn{1}{l|}{Home-grown}& \multicolumn{1}{l|}{CNN} & \multicolumn{1}{l|}{Layer}& \multicolumn{1}{l|}{A} & \multicolumn{1}{l|}{\checkmark} & \multicolumn{1}{l|}{\ding{53}} & \multicolumn{1}{l}{--} \\
    \multicolumn{1}{l|}{\cite{9786107}} & \multicolumn{1}{l|}{FINN\cite{umuroglu2017finn}}& \multicolumn{1}{l|}{BNN} & \multicolumn{1}{l|}{Layer}& \multicolumn{1}{l|}{A} & \multicolumn{1}{l|}{\checkmark} & \multicolumn{1}{l|}{\ding{53}} & \multicolumn{1}{l}{100} \\
    \hline
    \multicolumn{1}{l|}{Ours}  & \multicolumn{1}{l|}{NVDLA}& \multicolumn{1}{l|}{CNN}& \multicolumn{1}{l|}{Model} & \multicolumn{1}{l|}{A}& \multicolumn{1}{l|}{\checkmark} & \multicolumn{1}{l|}{\checkmark} & \multicolumn{1}{l}{1} \\
    \Xhline{1.5pt}
   \multicolumn{7}{l}{For Aim: A refers to architecture recovery while W refers to weight recovery.} \\
   \vspace{-30pt}
    \end{tabular}}
  \label{paper}%
\end{table}%

Although various side-channel techniques have been proposed to attack FPGA-based DNN accelerators, there is still a huge gap to apply them in practice. \textbf{(1) Simplified implementations}. A majority of works only target their homemade DNN accelerators \cite{yoshida2019model, li2021power, yu2020deepem,hua2018reverse,zhang2021stealing,9786107}, which are normally simplified, and easy to break. In contrast, full-fledged architectures in the real world usually involve more complex structure designs and optimizations, which complicate the side-channel analysis and attacks. \textbf{(2) Simplified models}. A number of works aim to steal Binary Neural Networks (BNNs) with binary values of model weights and activations \cite{yu2020deepem,9786107}. The feasibility of their extension to non-binarized DNNs is dubious. \textbf{(3) Simplified attack goals}. Many studies only attack individual network layers, but cannot achieve end-to-end extraction of the entire model \cite{tian2021remote,gupta2023ai,zhang2021stealing,9786107}. This restricts their practical values. Table \ref{paper} summarizes the comparisons of prior studies.

To address the above limitations, we propose \Name, 
a novel power side-channel attack to steal the architecture of DNN models on the practical NVDIA Deep Learning Accelerator (NVDLA). NVDLA serves as the standard way for DNN accelerator designs, and has been the mainstream implementation in many products. It includes the complex hardware-software stack, execution pipeline and runtime environment. Such designs cause the side-channel traces to be highly redundant and noisy, and exhibit no one-to-one relationship with the victim model. This significantly increases the attack difficulty (See Section \ref{sec:challenge} for more discussions).

The core concept of \Name is to model the side-channel extraction process as a sequence-to-sequence learning task.  We leverage powerful RNN-CTC and Transformer models to recover the architecture of the victim model from the power trace of its inference. Besides, we utilize the attention mechanism to localize the leakage point in the side-channel trace, which can shed light on the potential vulnerability of DNN accelerators and defense directions.

To our best knowledge, the only side-channel attack targeting NVDLA is \cite{gupta2023ai}. However, it has the following limitations: (1) it is not a remote attack and requires physical access to the victim device; (2) it requires the attacker to manually split the execution trace for different layers, and extract each layer individually; (3) it needs to train multiple learning models for each hyper-parameter. \Name can overcome these limitations: it can be launched remotely in the multi-tenant cloud context; it enables the adversary to automatically steal the model without any manual analysis or prior knowledge of the victim system. Besides, \Name is cost-efficient: the adversary only needs to train one end-to-end model, and run one inference process for extraction of the entire model, while prior attacks need thousands of rounds \cite{yoshida2019model, li2021power, yu2020deepem}.

Note that our attack goal is to steal the model architecture, which is the same as some previous works \cite{lou2021naspy,yan2020cache,hong20200wn,hu2020deepsniffer}. Stealing the architecture has high financial incentive as it is basis for building more valuable intellectual properties at incremental cost \cite{hong20200wn,lou2021naspy,lou2022ownership}. Besides, obtaining the architecture details can facilitate other attacks such as adversarial examples and membership inference. We provide two case studies in Section \ref{sec:case-study} to show how \Name can enhance these attacks. Some studies proposed methods to extract the model weights \cite{yoshida2019model,li2021power,yu2020deepem,hua2018reverse}. However, these attacks need to physically access the victim device or inject hardware trojans to collect more informative traces (Table \ref{paper}). How to remotely extract model weights is challenging and we leave it as a future work.

We perform extensive experiments to validate the effectiveness of \Name. Evaluation results show that  \Name can recover victim's model information with an error rate of 1\%. It is robust enough to resist certain levels of noise without a significant performance drop. 

\section{Background}
\label{sec:preliminary}
\subsection{Nvidia Deep Learning Accelerator (NVDLA)}
NVDLA is an open-source configurable architecture designed by Nvidia to accelerate deep learning inference. It is capable of computing convolution, activation, pooling, and normalization operations in the model inference. NVDLA can be configured as a large or small implementation, differing in the dimension of the cores and implementation of some specific engines (e.g., Rubik and DMA engines) \cite{nv_thesis}.



\begin{figure}[t]
	\centering
	\includegraphics[scale=0.3,trim=0 0 0 0]{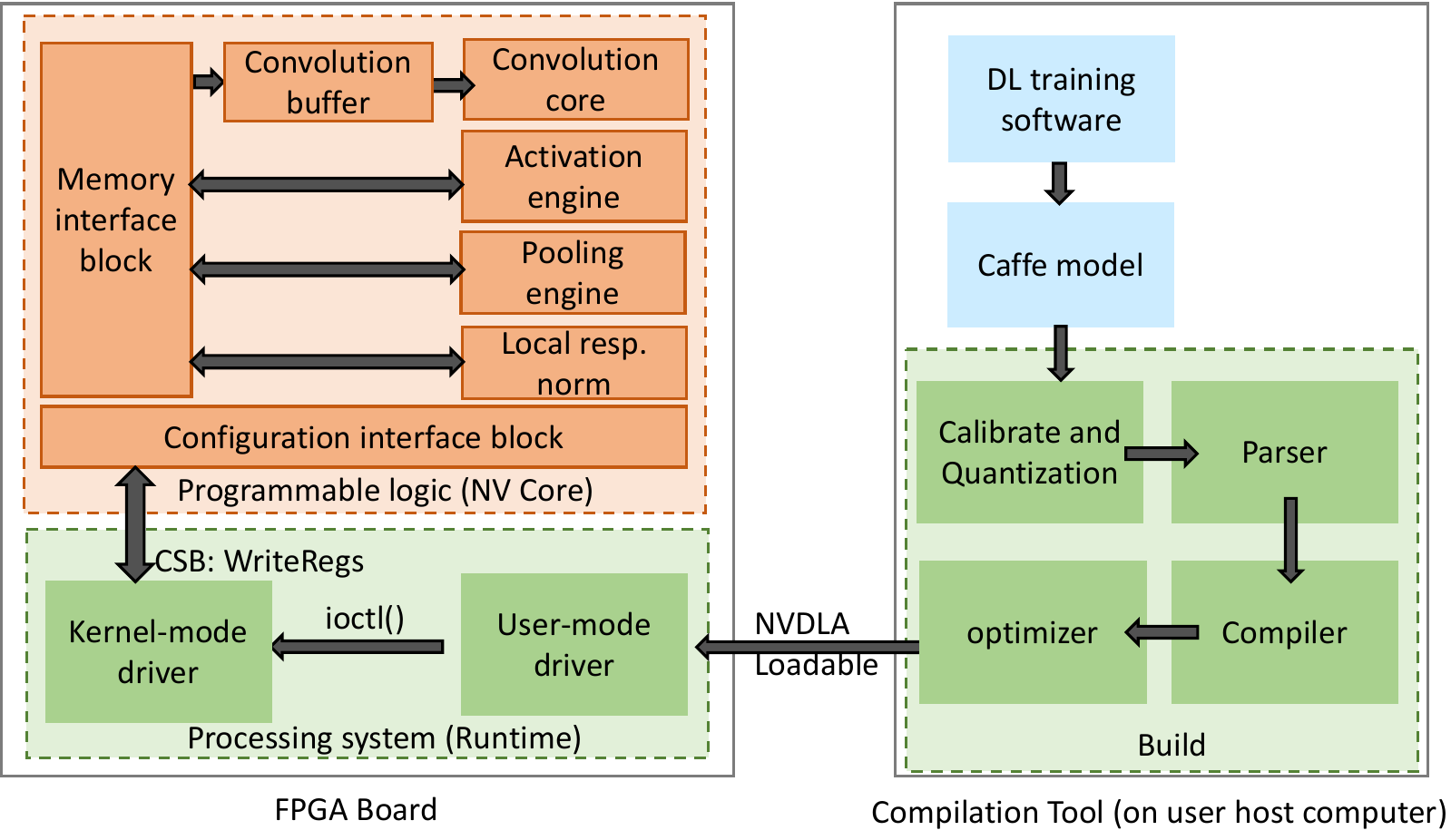}
    \vspace{-5pt}
	\caption{Architecture overview of NVDLA}
	\vspace{-10pt}
	\label{nv}
\end{figure}

Figure \ref{nv} shows the architecture overview of NVDLA. It can be divided into two parts, i.e. hardware design and software design. Hardware design is built as a series of pipeline stages containing various types of engines to regulate the behaviors of FPGA boards. Software design connects the users and hardware components, and is responsible for building and loading the DNN model for the FPGA to execute. 
The software design further consists of two components: (1) the compilation tools use the model pre-compiled by Caffe to generate a network of hardware layers supported by NVDLA, called loadable, which is calibrated by TensorRT.
(2) The runtime environment processes the calibrated loadable and runs it directly in the NVDLA environment. 

\subsection{Voltage Drop Sensor on FPGA}
All the components on an FPGA chip share one power distribution network (PDN). Intensive switching activities may cause voltage fluctuations in the PDN. The PDN can be modeled as an RLC circuit, where a resistor (R), an inductor (L), and a capacitor (C) are connected in series or in parallel. Therefore, the transient voltage drop seen by a circuit can be modeled as \cite{zhao2018fpga}: $V_{drop}=IR+L\frac{di}{dt}$, 
where the transient response term $ L\frac{di}{dt} $ reflects the intensity of the switching activity on the FPGA. In typical CMOS
circuits, combinational logic delays can be modeled to be
inversely proportional to the voltage supplied to each gate \cite{pant2008design}. Hence, the information of the switching activities can be inferred from the logical delay. 

In this paper, we use a  time-to-digital converter (TDC) to read the combinational logic delay, where a clock signal propagates through a chain of buffers as the voltage drop sensor.  
As there are discrepancies in switching activities for different types of calculation on other parts of the FPGA, the voltage drop value may be different, which leads to different delay measurements in TDC. These different delays can cause different propagation lengths in the delay line, resulting in different values in the latches. Therefore, the activities of other circuits can be identified through the readout of the TDC. This has been demonstrated in various studies \cite{tian2021remote, moini2021remote,gravellier2021remote}.

\subsection{Profiled Side-Channel Attacks}
Profiled side-channel attacks are one of the most powerful attacks \cite{standaert2009compare}. The adversary generates a profile
through a similar or the same device of the target, and computes the secret information by matching the profile with the victim's execution trace. A profiled
side-channel attack can be divided into two phases. First, a profiling phase allows the adversary to characterize its physical leakages when running the target application. Suppose the adversary has an input secret set $ \mathbf{s}=\{s_1,...,s_n\} $. He obtains $N$ side-channel traces $ \mathbf{{T_{i,n}}} $ for each input $ s_i $, and builds the mapping $ \mathit{f}: \mathbf{T_{i,n}}\mapsto s_i $. 
This mapping can be learned through template creating and machine learning methods. Second, an exploitation phase is launched to perform secret recovery by profile matching. The adversary performs the side-channel attack using the mapping \textit{f}. With additional \textit{q} traces $ {T}'_1,...,{T}'_q $ collected from the device under attack, the secret $ s^{'}_i $ can be guessed as $ s^{'}_i=\mathit{f}({T}'_i) $.  

In this paper, we adopt machine learning for profile learning and matching to attack FPGA-based DNN accelerators. It has several advantages. First, the adversary is able to extract information even if there are no visible patterns in the power trace. As most calculations on FPGAs are done in parallel, manually analyzing the pattern may be futile. Second, the adversary can utilize all information in a single power trace, as machine learning can effectively handle high-dimensional data. In contrast, traditional methods require the selection of points of interest (POI) to narrow down the information for the attack. Besides, machine learning models have higher resilience against the noise in the side-channel trace than conventional statistical methods. 


\subsection{Sequence-to-sequence (seq2seq) Learning}

Seq2seq learning achieves state-of-the-art prediction accuracy in various tasks like speech recognition \cite{amodei2016deep}, machine translation \cite{neubig2017neural}, image captioning \cite{islam2019sequence}, question answering \cite{palasundram2020enhancements}, etc. Since side-channel power traces are essentially sequential data, seq2seq learning is a natural fit for analyzing such leaking patterns. However, only very few works \cite{hu2020deepsniffer, lou2021naspy} have applied seq2seq learning to side-channel analysis, and none of them realized automated attacks on FPGA-based accelerators, which have much more noise and more complicated monitoring architecture.  

Commonly used architectures for seq2seq learning includes Transformer \cite{vaswani2017attention}, Connectionist Temporal Classification (CTC), and Recurrent Neural Network (RNN). 
\Name adopts two models, i.e, Transformer and RNN-CTC, for side-channel extraction. 
In the RNN-CTC model, the output of the RNN is passed into the CTC decoder. Aligning the operation sequence with the variable-length power sequence presents a challenge. To address this, the CTC decoder introduces a "blank" label $\epsilon$ that serves no specific correspondence and can be easily excluded from the output. The Transformer model consists of an encoder and a decoder. 
The input sequence is transformed by the encoder, generating an abstract representation that captures the learned features. Subsequently, the decoder utilizes this abstract representation to predict the subsequent output step-by-step, building upon the previous output. To enhance feature learning from the event sequence, a convolution layer is introduced before the encoder.

\section{Attack Overview}
\label{sec:overview}

\subsection{Threat Model}
We follow the \textit{same threat model} of remote power side-channel attacks in previous works \cite{tian2021remote,moini2021remote,luo2021deepstrike}: the adversary and victim's accelerator share the same FPGA board (but logically separated) with the same PDN, which can be realized in the multi-tenant FPGA-based cloud. The adversary is able to deploy and fully control his own malicious circuits. However, he cannot physically access the FPGA board or control any part of the victim's circuit. He can only deduce the victim's activity from his own circuit. 

\begin{figure}[t]
	\centering
	\includegraphics[width=\linewidth]{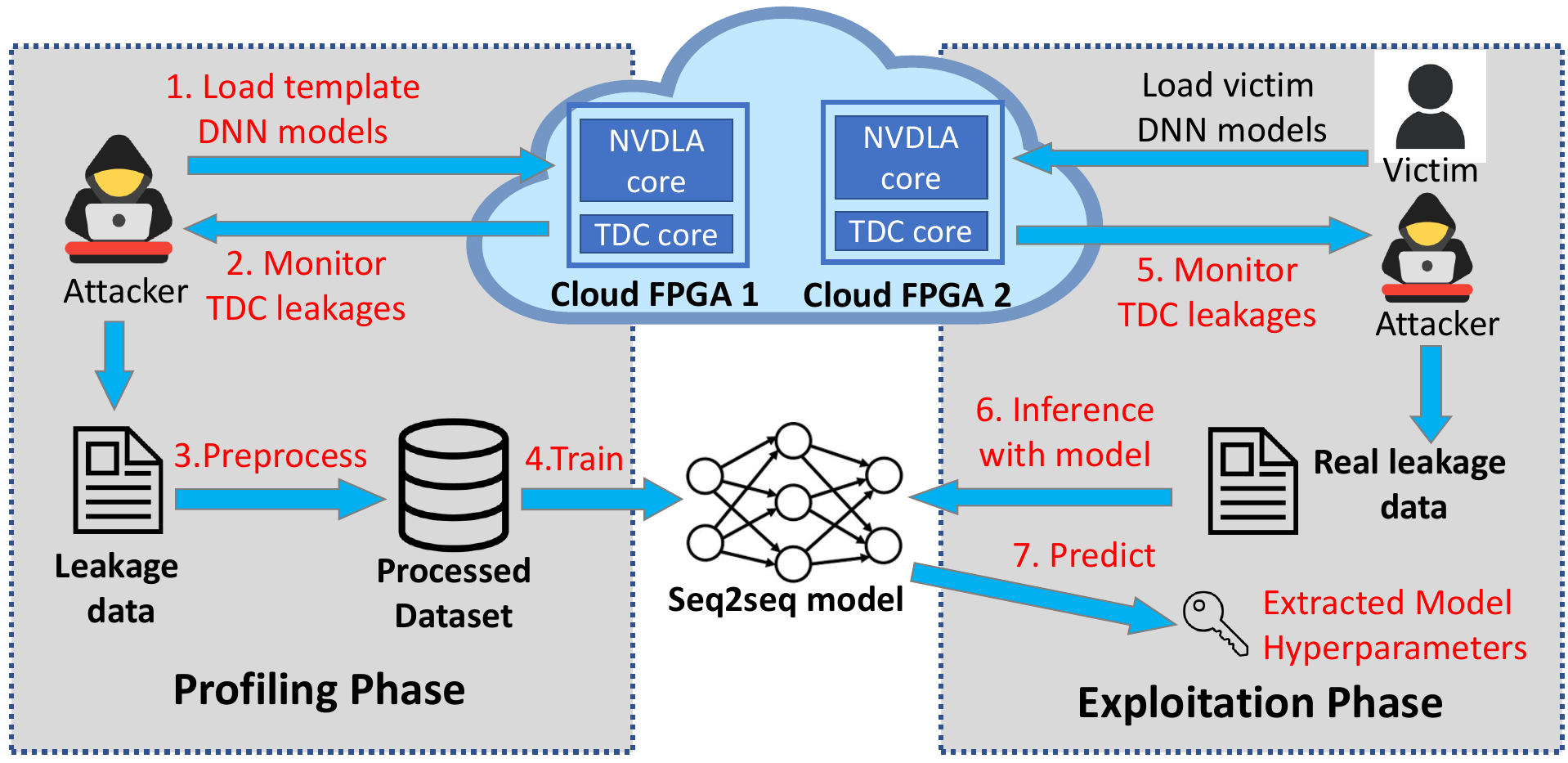}
	\caption{Attack workflow in \Name}
    \vspace{-15pt}
	\label{md}
	
\end{figure}

Figure \ref{md} illustrates the workflow of \Name, which consists of two phases. (1) In the \textit{profiling} phase, the adversary deploys various types of DNN models on a cloud FPGA board and collects the corresponding power traces with the TDC sensor. With the profiling information he can establish the seq2seq model $f$ that predicts the model architecture from the power trace. (2) In the \textit{exploitation} phase, the adversary deploys the TDC circuit on the same FPGA as the victim's model. He only needs to remotely collect the readout of this sensor for \textit{one} inference process of the target model. Then he can use $f$ to extract the victim model's architecture details.


\subsection{Challenges of Attack Design}
\label{sec:challenge}
The NVDLA implementation raises the following challenges for designing remote side-channel attacks. 


\textbf{(1) NVDLA has complicated hardware and software architecture.}
From the software perspective, pre-compiled neural networks are first loaded by the user space runtime driver and then submitted to the kernel mode driver for the subsequent inference. The kernel mode driver schedules layer operations on NVDLA and programs the NVDLA registers to configure each functional block. From the hardware perspective, functional computation blocks are assembled as a pipeline to run submitted tasks, and the interrupt signal is asserted when the task is completed. Due to such complex hardware-software co-design, the captured power trace from NVDLA does not directly reflect the information of each layer and the data inside. It is mixed with other hardware-level power information on the FPGA, as well as the software-level power information related to CPU execution. This is much more complex than other simple accelerator designs, and significantly increases the attack difficulty. 

\textbf{(2) NVDLA has a parallel design to boost the performance.}
All functional blocks in NVDLA have duplicated register groups known as ping-pong buffers. Configurations of the next layer will be transmitted to the other register while running the current layer \cite{nv}. Due to this parallel design, NVDLA data flow may not have a strict one-to-one relationship with the layers of a model. These settings make the model extraction attack much more challenging.
 
\textbf{(3) The collected side-channel data are redundant and noisy.}
A side-channel trace of an inference process can contain an extremely large amount of data points (more than 300K) due to the high monitoring frequency, which places heavy demands on the underlying compute infrastructure. Such a scale of data will induce great pressure on the data processing, making it infeasible to perform manual analysis as adopted in existing works. 
Moreover, the raw side-channel data may exhibit random noise originating from the hardware activities, a consequence of the intricate system optimization and runtime dynamics. This noise has the potential to substantially diminish the accuracy of the extraction process.

\section{Design Details}
\label{sec:detail}


To address the above challenges, we introduce innovative designs for the power monitor, dataset processing and seq2seq model. The detailed mechanisms are elaborated below.

\subsection{TDC-based Power Monitor}
\begin{figure}[t]
	\centering
	\includegraphics[width=0.9\linewidth,height=5cm]{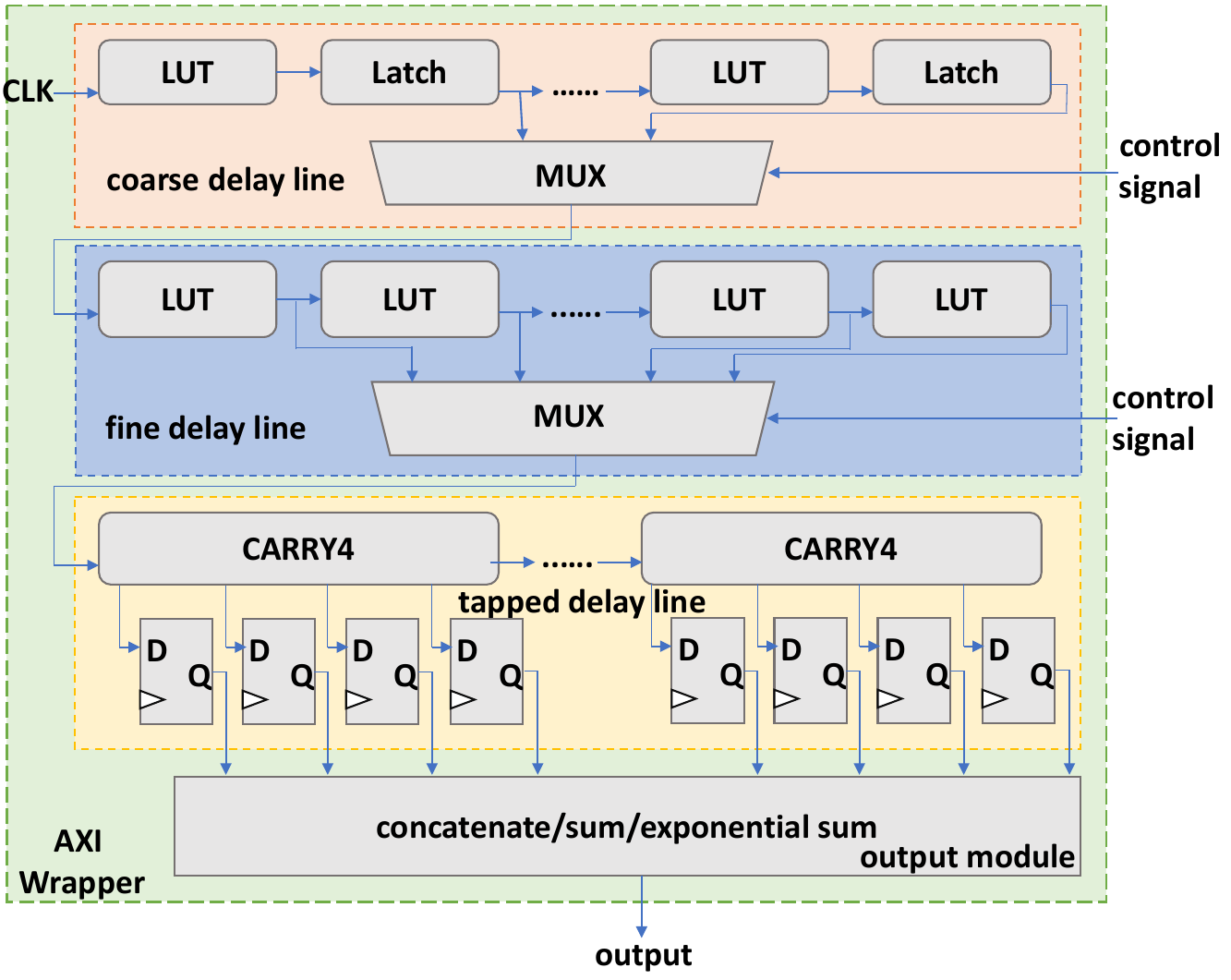}
	\caption{Architecture details of TDC}
	\vspace{-15pt}
	\label{tdc-detail}
	
\end{figure}
\Name utilizes TDC readouts to infer power information from the inference process. Figure \ref{tdc-detail} illustrates the architecture details of the TDC, where the clock signal is fed into an adjustable coarse delay line and fine delay line to obtain an initial delay, which is then sent to a tapped delay line. The initial delay can be dynamically configured by controlling the multiplexer (MUX) to modify the number of logic elements forming the coarse and fine delay lines during the calibration process, thus altering the delay duration. To ensure the TDC's functionality is not optimized during synthesis or implementation phases, the \verb!DONT_TOUCH! attribute is set for the coarse delay line, fine delay line, and tapped delay line.

The coarse delay line consists of replicated look-up table (LUT) and latch modules to provide a significant amount of delay. In contrast, the fine delay line is equipped with replicated LUT modules to offer smaller delay. The tapped delay line, which employs carry chains, is composed of \verb!CARRY4! primitives, with their \verb!CO! outputs registered by four dedicated D flip-flops. During each readout, it counts the taps which the clock signal has reached and gives a raw value. The output can then be concatenated or converted into a sum or exponential sum, depending on the configuration set in the TDC IP settings. 
The TDC output is routed to the ARM processor via AXI-4 buses. We develop a C-based driver program running on the ARM processor to read the TDC output using the \verb!mmap! system call on the addresses specified in the Vivado IP Integrator.

\begin{figure}[t]
	\centering
	\includegraphics[width=0.9\linewidth,height=4.5cm]{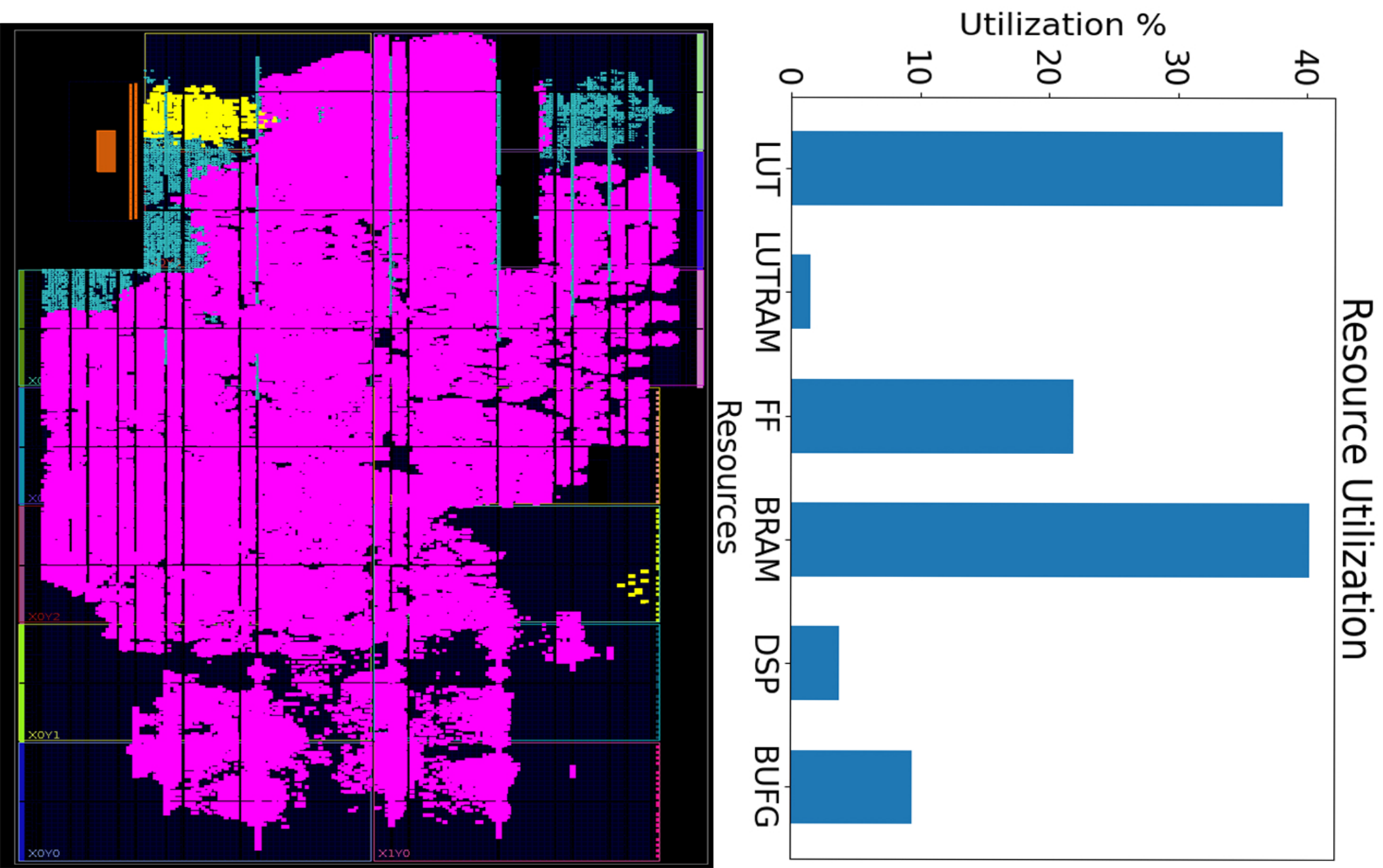}
	\caption{(Left) The floorplan showing the location of NVDLA (in purple) and the TDC sensor (in yellow) on the FPGA. (Right) Resource utilization on FPGA}
	\vspace{-10pt}
	\label{floorplan}
	
\end{figure}

Note that it is important to perform TDC calibration, i.e., adjusting its initial delay, prior to output measurement. We implement the calibration process as two loops in our driver. We iteratively test all combinations of possible values for the fine and coarse delay line lengths to select the optimal initial delay value. This ensures that the Hamming weight in the tapped delay line of the TDC is half the length of the tapped delay line, providing room for power measurements to increase or decrease the Hamming weight. Although prior works \cite{0Understanding,gravellier2021remote} highlight the importance of TDC placement and routing constraints for obtaining useful side-channel information, in practice, the adversary may not have the privilege of determining such configurations. In \Name, we do not need to manually set the TDC location, yet still successfully initiate the attack. 
Figure \ref{floorplan} shows the floorplan of our implemented design, with the resource utilization on the FPGA.

\begin{figure}[t]
	\centering
	\includegraphics[scale=0.26,trim=0 0 0 0]{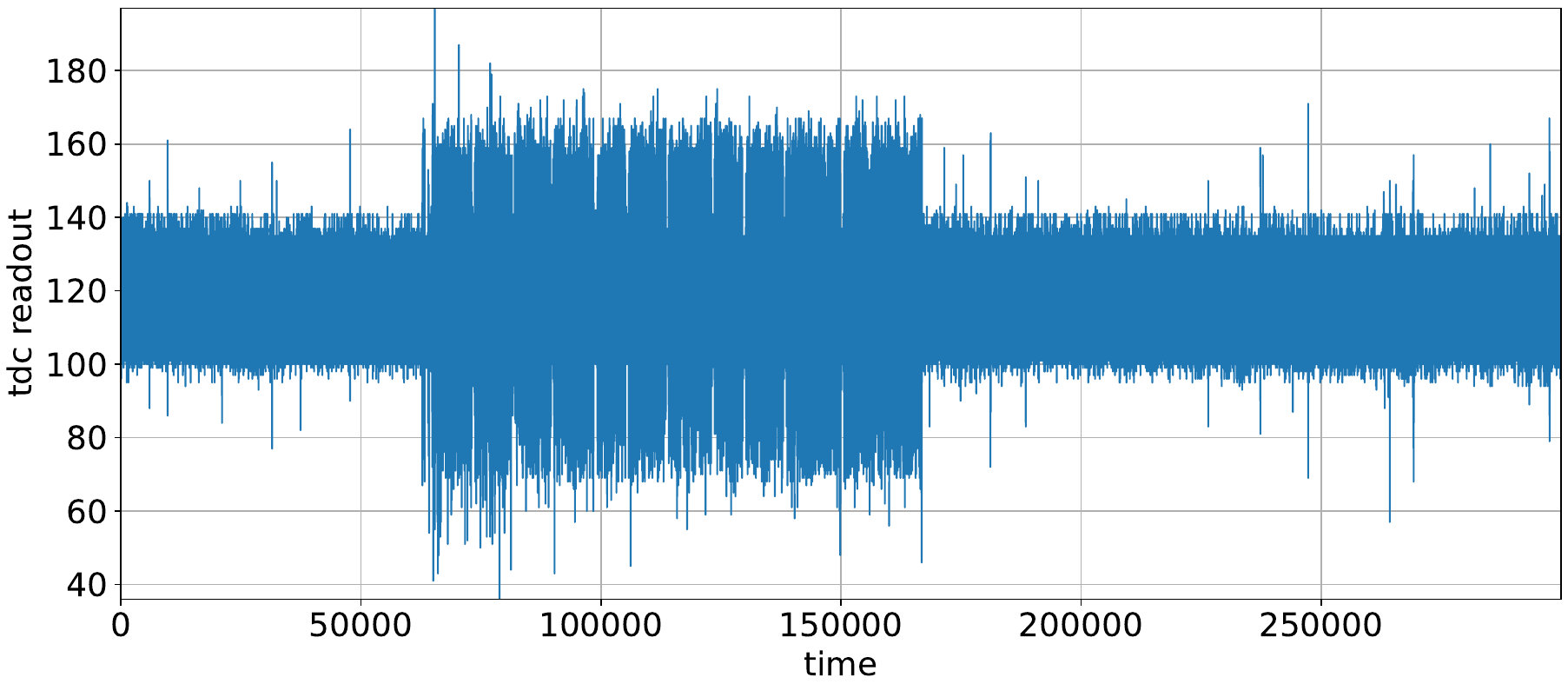}
	\caption{TDC readouts for one inference process}
	\vspace{-15pt}
	\label{cap}
	
\end{figure}

\begin{figure*}[t]
	\centering
	\includegraphics[width=0.85\linewidth]{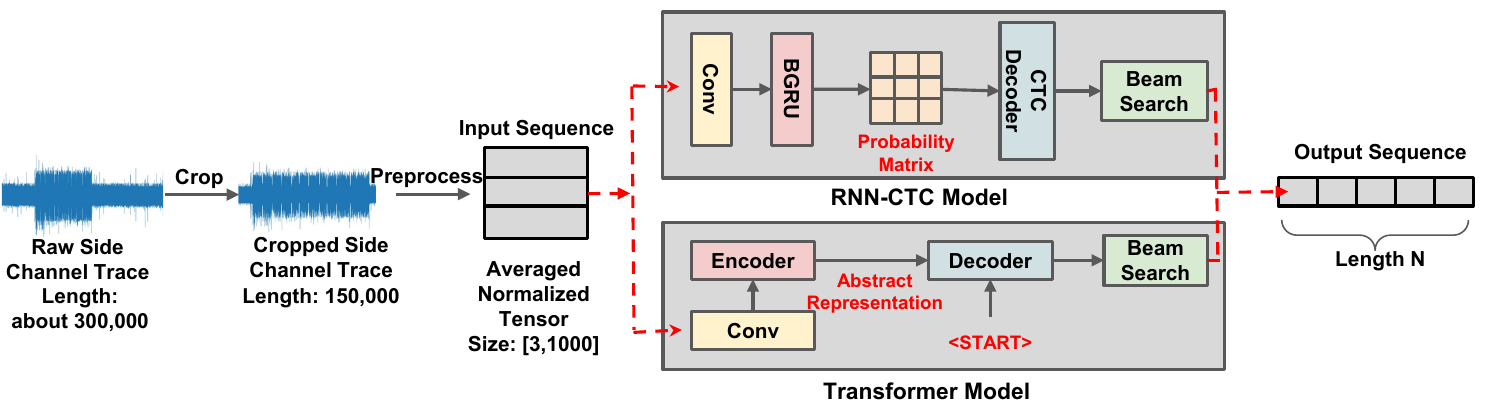}
	\caption{Overview of two alternative attack models}
	\vspace{-20pt}
	\label{network}
\end{figure*}

\subsection{Dataset Formulation and Preprocessing}
To build a generalized seq2seq model for side-channel extraction, we need to collect a dataset that covers different types of DNN models and operations in the profiling phase. To achieve this, we generate 160 different random models and deploy them on NVDLA. These models have random numbers (in the range of [2, 16]) and types of network layers.  They include 12 convolution layers (with the kernel size of 2, 3, 4, 5 and output size of 10, 20, 30), 4 pooling layers (with the kernel size of 2, 3, 4, 5), 5 fully-connected layers (with the output size of 100, 200, 300, 400, 500), 1 relu layer, and 1 softmax layer. These models are pre-trained by Caffe, and then calibrated by TensorRT and compiled by the NVDLA compiler. They are generated on the host computer, and then executed by NVDLA runtime on the FPGA.

The power trace captured by TDC is a sequence in which the TDC readouts are arranged in the temporal order. They represent the switching activities on the FPGA at different moments. To better control the data collection process, we use a bash script on the board to start the TDC measurement program concurrently with the model inference on NVDLA, and terminate it when the inference task is completed.

Figure \ref{cap} shows an example of the TDC readouts for one inference. The $x$-axis denotes the time while the $y$-axis represents the corresponding power consumption value. The middle part of this trace has larger fluctuations, indicating the power-related information leakage from NVDLA. Therefore, we crop the trace and only use the data within this effective period with the time coordinate range [50,000, 200,000] for all traces. Training the model with the full trace also yields successful attack results but requires longer training time. For each model, we collect approximately 2,700 traces as the training set, and 200 traces as the test set. 
As every single trace has a large amount of data points which may decrease the training efficiency, we reshape it to a 2D matrix with the size of $3\times50,000$. Then we compute the average of 50 data points, normalized them by subtracting the mean and divided by the standard deviation, for efficient model training. 

We label each power trace with a sequence of types for each layer. Given that convolutions take a majority in neural networks, we differentiate different kinds of convolution layers with different labels. We also set four labels for the pooling layer, fully-connected layer, relu layer, and softmax layer, respectively. Table \ref{label} shows the label for each type of network layer. 
As required by the CTC decoder (Section \ref{sec:model}), there is also one label representing a blank operation. Therefore, we have 17 different possible labels in total. The combination of these labels forms the label sequences.

\begin{table}[h]
  \centering
  \caption{Prediction labels for different types of layers. For example, the label sequence for model with one 5*5 conv layer (10 output channels) and one fc layer is represented as [9,13].}
    \begin{tabular}{c|c}
    \Xhline{1.5pt}
    \textbf{Layer} & \multicolumn{1}{l}{\textbf{Label}} \\
    \Xhline{1.5pt}
    conv layer, kernel size 2*2, output channel 10, 20, 30 & 0-2 \\
    conv layer, kernel size 3*3, output channel 10, 20, 30 &  3-5\\
    conv layer, kernel size 4*4, output channel 10, 20, 30 &  6-8\\
    conv layer, kernel size 5*5, output channel 10, 20, 30 &  9-11\\
    pooling layer & 12 \\
    fully-connected layer & 13 \\
    relu layer & 14 \\
    softmax layer & 15 \\
    \Xhline{1.5pt}
    \end{tabular}%
  \label{label}%
\vspace{-15pt}
\end{table}%

\subsection{Seq2seq Model}
\label{sec:model}


\Name adopts two \textit{alternative} seq2seq learning models to extract the DNN architecture from the side-channel trace, as shown in Figure \ref{network}. The first one is RNN-CTC. This model uses some convolution layers to extract features from the input sequences, and then 2 RNN layers to propagate the information.
To enhance long-term memory capabilities, the RNN module adopts the bidirectional gated recurrent unit (BiGRU).
The DNN models in the profiling phase are randomly generated and may not have strong relationships between the layers as they do in the actual functional model design. Hence using RNN may not give the best results. However, we believe it can have better performance in the actual situation as the relation of layers can be utilized. 
The RNN layer produces a probability distribution for each input, which is subsequently passed into the CTC decoder.
The number of training parameters is 564,590.
Despite the challenge of aligning operation sequences with varying lengths to their corresponding power sequences, the CTC decoder overcomes this hurdle by introducing a "blank" label. Ultimately, the output sequence with the highest prediction probability is determined using beam search.
To obtain better training results, Adam optimization and the OneCycleLR scheduler are also used in our model.

The second one is the Transformer model, which adopts a single-layer encoder and single-layer decoder. This model utilizes weight sharing between the decoder embedding and the decoder projection to enhance generalization. It is also trained using the Adam optimizer with OneCycleLR scheduler. The performance is evaluated by the cross-entropy loss function.

The Transformer model can also localize the leakage point in the TDC trace with the attention mechanism, helping us better understand the side-channel vulnerabilities of the DNN inference. Attention is a powerful technique that allows the model to selectively focus on specific parts of the input when making predictions. It computes a weighted sum of input features with the weights representing the importance or relevance of each feature to the current prediction. We can leverage the multi-head attention in the Transformer model to identify the high attention weights, which correspond to the potential leakage points. See Section \ref{sec:leakage-localization} for more analysis.

\begin{figure*}
  \centering
		\subfloat[1 CNN Layer\label{fig:a}]{
			\includegraphics[scale=0.38]{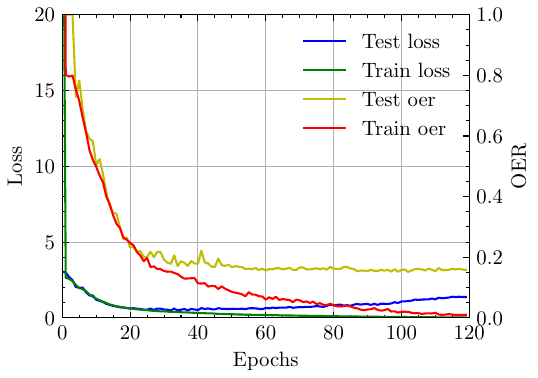}}
		\subfloat[2 CNN Layers\label{fig:c}]{
			\includegraphics[scale=0.38]{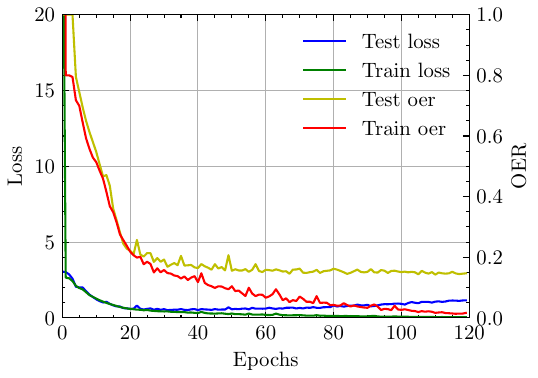}}
		\subfloat[3 CNN Layers\label{fig:e}]{
			\includegraphics[scale=0.38]{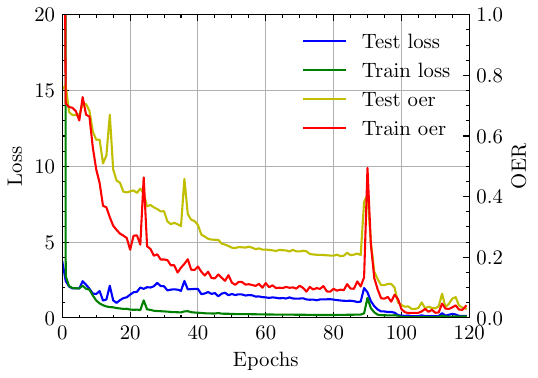}}
		\subfloat[4 CNN Layers\label{fig:b}]{
			\includegraphics[scale=0.38]{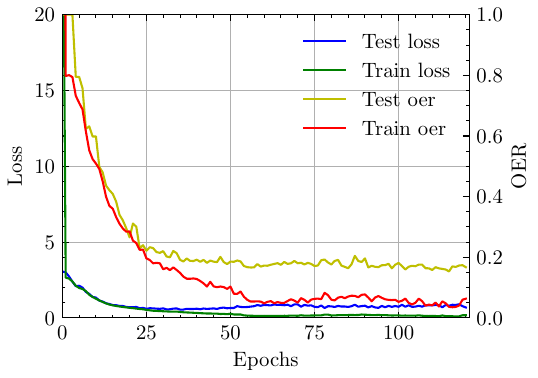} }
		\subfloat[5 CNN Layers\label{fig:d}]{
			\includegraphics[scale=0.38]{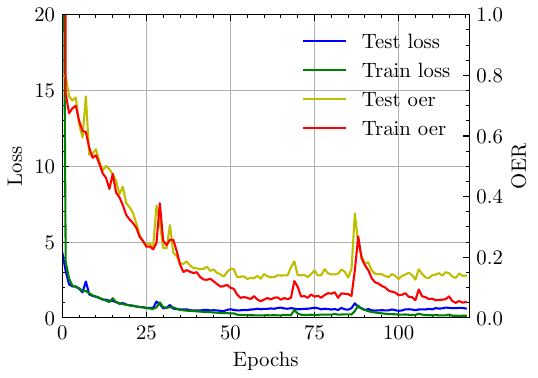}
			}
\caption{Loss and OER trends when training the RNN-CTC model with different numbers of CNN layers.}
\vspace{-10pt}
\label{oer-loss}
\end{figure*}

\section{Evaluation}
\label{sec:evaluation}

We adopt the Xilinx Zynq-7000 SoC ZC706 board (xc7z045ffg900-2) as our testbed. Due to the restriction of our hardware, we adopt the small implementation of NVDLA. \Name can be generalized to the large implementation as they perform the same operations \cite{nv}. The ARM processor of the board runs Ubuntu 16.04 OS, which supports NVDLA and the TDC driver. Vivado 2019.1 is used to design the hardware. The clock frequency is 10MHz for NVDLA and 150MHz for TDC. The board sends the TDC readouts to the host computer through Ethernet with the \verb!scp! command. Pytorch (1.13) and CUDA (11.6) are adopted to train models running on the server with a Nvidia GeForce RTX 3090 GPU. The training process takes approximately 6 GPU hours.

We utilize two metrics to evaluate the attack effectiveness. First, as the model architecture is represented as a sequence with each element representing the type of the layer, we adopt Operation Error Rate (OER) \cite{lou2021naspy} to measure the prediction accuracy. 
It is calculated as $ OER = L({s}',s)/\left \| s \right \| $, where $ \left \| s \right \| $ is the sequence length of $ s $, and $ L({s}',s) $ is the edit distance (Levenshtein) between the ground-truth sequence $ s $ and predicted operation sequence ${s}'$. A smaller OER indicates higher accuracy. Second, we also use loss functions. In the RNN and Transformer models, we use the CTC loss and cross-entropy loss for evaluation, respectively. 

We find it is difficult to compare \Name with prior attacks. As shown in Table \ref{paper}, most works fall into the category of physical side-channel attacks and require to set up physical instruments to obtain the power traces. For remote attacks \cite{tian2021remote,zhang2021stealing,9786107}, they only target layer extraction and cannot achieve end-to-end model stealing. Due to the distinct settings and attack goals, we mainly report the attack results of \Name.


\subsection{Model Extraction Results}

\noindent\textbf{Model performance.}
We employ our collected dataset to train both the RNN-CTC and Transformer models. 
For the RNN-CTC model, we utilize 3 CNN layers with an RNN dimension of 128, and train the model for 120 epochs. For the Transformer model, we incorporate one encoder layer and one decoder layer. 
The model input and output dimension, denoted as $d_{model}$, is configured as 256. We incorporate 8 parallel attention layers, with projected queries, keys, and values having dimensions of $d_k = d_v = d_{model}/h$, where $h$ represents the number of attention heads. The positional encoding and optimizer settings adhere to the implementation outlined in \cite{vaswani2017attention}. With these configurations, we train the model for 100 epochs.

We run inference for all randomly generated models in NVDLA on both MNIST and CIFAR-10 datasets. For MNIST, using RNN-CTC, we achieve OER of 0.01 and 0.02 on the training and test set, respectively; using Transformer, we get OER of 0.02 and 0.15 on the two sets. For CIFAR-10, with RNN-CTC, we achieve OER of 0.04 and 0.16 on the training and test set, respectively; with Transformer, we get OER of 0.10 and 0.15 on the two sets. We observe that both models exhibit strong performance on the training dataset. RNN-CTC outperforms Transformer on the test set, mainly because its stronger ability to handle variable-length input and output sequences. However, Transformer is more interpretable in analyzing the leakage points in the trace (Section \ref{sec:leakage-localization}). Below we mainly use RNN-CTC on MNIST for more evaluation. 

\begin{table}
\begin{subtable}[c]{0.48\linewidth}
\centering
    \resizebox{\linewidth}{!}{
    \begin{tabular}{c|c|c|c|c}
    \Xhline{1.5pt}
    \multicolumn{1}{c|}{\textbf{\# of}} &     \multicolumn{2}{c|}{\textbf{Best Loss}} &     \multicolumn{2}{c}{\textbf{Best OER}} \\ \cline{2-5}
    \textbf{layers} & \textbf{Train} & \textbf{Test} & \textbf{Train} & \textbf{Test} \\
    \Xhline{1.5pt}
    1     & 0.02 &  0.49 & 0.01 & 0.15 \\
    2     & 0.04 & 0.51 & 0.01 & 0.14 \\
    \textbf{3}     & \textbf{0.04} & \textbf{0.10} & \textbf{0.01} & \textbf{0.02} \\
    4     & 0.09 & 0.54 & 0.04 & 0.15 \\
    5     & 0.13 & 0.44 & 0.05 & 0.12 \\
    \Xhline{1.5pt}
    \end{tabular}}
\subcaption{Number of CNN layers}
\label{tab1}
\end{subtable}
\begin{subtable}[c]{0.48\linewidth}
\centering
    \resizebox{\linewidth}{!}{
    \begin{tabular}{c|c|c|c|c}
    \Xhline{1.5pt}
    \multicolumn{1}{c|}{\textbf{RNN}} &     \multicolumn{2}{c|}{\textbf{Best Loss}} &     \multicolumn{2}{c}{\textbf{Best OER}} \\ \cline{2-5}
    \textbf{dims} & \textbf{Train} & \textbf{Test} & \textbf{Train} & \textbf{Test} \\
    \Xhline{1.5pt}
    64     & 0.40& 0.50 &0.14 &0.16 \\
    96     & 0.33& 0.62 &0.10 &0.17 \\
    \textbf{128}     & \textbf{0.04} & \textbf{0.10} & \textbf{0.01} & \textbf{0.02} \\
    256     &0.06 &0.51  &0.02 &0.15 \\
    512     &0.03 &1.45  &0.01 &0.19 \\
    \Xhline{1.5pt}
    \end{tabular}}
\subcaption{RNN dimensions}
\label{rnn}
\end{subtable}
\caption{Impact of RNN-CTC configurations}
\vspace{-15pt}
\end{table}

\noindent\textbf{Ablation studies.}
We further investigate the impact of hyper-parameters on the RNN-CTC model. We first consider different numbers of CNN layers used for feature extraction (1 $\sim$ 5). Figure \ref{oer-loss} shows the trends of the training and test losses and OER when training the RNN-CTC model. We observe that OER and loss in all figures drop when the training proceeds. Table \ref{tab1} reports the best OER and loss values for different numbers of convolution layers. We find that the RNN-CTC model with 3 convolution layers gives the best performance. 
We also test the prediction accuracy of the RNN-CTC model with different RNN dimensions. Table \ref{rnn} reports the best loss and OER when the RNN dimension is set as 64, 96, 128, 256, and 512 respectively. We observe that 
RNN with the dimension of 128 has the best performance. We will adopt such optimal hyper-parameters in the following experiments.  

\subsection{Robustness Analysis}
\noindent\textbf{Side-channel noise}.
We first evaluate the robustness of the prediction model against side-channel noise. We add different amounts of Gaussian noise to the side-channel trace, and measure to what extent the extraction accuracy will be affected by such noise. The scale of the injected noise is calculated as $P_n=P_s/(10^{SNR/10})$, where $P_s$ is the power of the input sequence, and SNR is the signal-to-noise ratio (SNR), defined as the power ratio of the input to the noise in decibels. 

\begin{table}[htbp]
\vspace{-10pt}
  \centering
  \caption{Accuracy with different SNRs}
    \begin{tabular}{c|c|c}
    \Xhline{1.5pt}
    \multicolumn{1}{c|}{\textbf{SNR(dB)}} & \multicolumn{1}{c|}{\textbf{OER}} & \multicolumn{1}{c}{\textbf{Loss}} \\
    \Xhline{1.5pt}
    No noise & 0.05  & 0.18 \\
    50    & 0.05      & 0.18 \\
    40    & 0.09      & 0.35 \\
    30    & 0.20      & 0.93 \\
    25    & 0.34      & 1.34 \\
    10    & 0.71  & 2.83 \\
    \Xhline{1.5pt}
    \end{tabular}
  \label{snr}%
\end{table}%
Table \ref{snr} reports the extraction results under various scales of noise. We observe that RNN-CTC has no accuracy drop when the SNR is above 50dB. The side-channel noise only has a minor effect on the model when the SNR is 40dB. Considering that our input sequence collected from TDC already contains a certain level of noise, the actual threshold of SNR that can preserve the extraction accuracy should be even lower than the tested value. Therefore, our model has high robustness to resist the noise in the measured trace. 

\noindent\textbf{Placement location of TDC.}
Next we explore the effect of the TDC locations on the board during model extraction. Prior studies \cite{0Understanding, gravellier2021remote} showed that the outputs of the TDC are highly sensitive to its location. Hence, it is essential for the adversary to find the optimal place to implement the TDC. To evaluate this factor, we place the TDC in 5 different locations: top-left, bottom-left, center, top-right, and bottom-right. 
We set the location constraints using Pblock in Vivado. For each location, we collect 600 TDC traces and then re-evaluate the prediction of both models, as shown in Table \ref{tdcloc} (the ``w/o.'' columns). We observe that a different location of TDC can indeed degrade the extraction accuracy, as the adversary's side-channel power traces during training and inference is different. 


\begin{table}[t]
  \centering
  \caption{Sensitivity to TDC locations}
    \begin{tabular}{c|c|c|c|c}
    \Xhline{1.5pt}
    \multirow{2}{*}{\textbf{TDC location}} & \multicolumn{2}{c|}{\textbf{OER (RNN)}} & \multicolumn{2}{c}{\textbf{OER (Transformer)}} \\ \cline{2-5}
    & w/o. & w/. & w/o. & w/. \\
    \Xhline{1.5pt}
    Original & \multicolumn{2}{c|}{0.14} & \multicolumn{2}{c}{0.15} \\ \hline
    Center & 0.51 & 0.15 & 0.59 & 0.11 \\
    Top-right & 0.4 & 0.23  & 0.49 & 0.23  \\
    Top-left & 0.39 & 0.20 & 0.43 & 0.20 \\
    Bottom-right & 0.25 & 0.23 & 0.45 & 0.26 \\
    Bottom-left & 0.41 & 0.19 & 0.59 & 0.14 \\
    \Xhline{1.5pt}
    \end{tabular}
  \label{tdcloc}%
\vspace{-10pt}
\end{table}

In the real-world multi-tenant cloud scenario, the adversary may not have the permission to select the location for implementing his TDC, which is allocated by the cloud provider. To bridge this gap and make our attack practical, the adversary can augment his training dataset by placing the TDC in multiple locations and collecting the comprehensive power traces in the offline phase. Then the prediction model will be more general and robust against TDC placement. We train such a model with an augmented dataset which includes additional 10\% of TDC trace for each of the above 5 locations. Table \ref{tdcloc} (the ``w/.'' columns) shows that the prediction errors are significantly reduced, and close to that of the original location.

\subsection{Leakage Point Localization}
\label{sec:leakage-localization}
\begin{figure}[t]
	\centering
	\includegraphics[scale=0.25,trim=0 0 0 0]{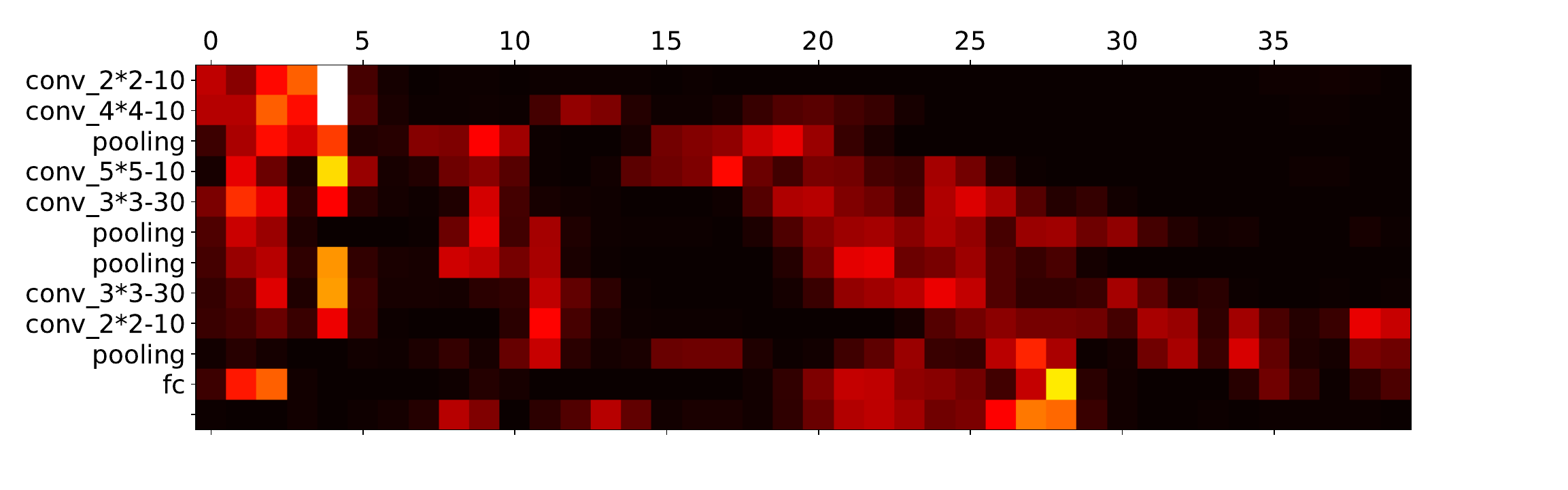}
     \vspace{-5pt}
	\caption{Attention weight (brighter color means larger weights)}
	\vspace{-15pt}
	\label{attn2}
\end{figure}

We show how to use the Transformer model to localize the leakage point in the power trace. 
As described in Section \ref{sec:model}, we compute the attention weight as the indicator of leakage. Given that the input trace is much longer than the output sequence, the attention weight matrix is too long to be easily illustrated. Therefore, we average every 25 attention weights. Figure \ref{attn2} shows an example of the attention weights for one TDC trace. The model architecture is shown on the left side, where ``fc'' represents a fully connected layer and ``conv\_2*2-10'' represents a convolution layer with the kernel size of $2\times2$ and output feature map dimension of 10. The number at the top represents the input sequence.



From Figure \ref{attn2}, we observe that the highest attention weights normally appear very early in the trace, which indicates that major leakage of model information is from the beginning. Considering that NVDLA sends the layer-related information to the relevant register from ARM to FPGA before computation, we hypothesize that the major leakage may occur during the transmission or storage of this information. This suggests the importance of providing more protections over these particular locations as the countermeasure. We leave such interpretability-based defenses as future work.

\section{Case Studies}
\label{sec:case-study}

As mentioned in Section \ref{sec:intro}, extracting model architectures can not only compromise model intellectual property, but also facilitate some attacks to deep learning. We present two case studies to demonstrate how \Name can enhance the adversarial examples and membership inference attack.

\subsection{Enhancing Adversarial Examples}
A popular security threat to deep learning models is adversarial examples (AEs), which are created by adding human-invisible perturbations to normal samples to mislead the victim model \cite{szegedy2013intriguing,goodfellow2014explaining}. Over the years, numerous attack methodologies have been proposed to generate effective AEs, which can be classified into two categories based on the threat model. The first one is \textit{white-box} attacks. The adversary has knowledge of the model parameters, based on which he precisely crafts the adversarial perturbations. Typical methods include FGSM \cite{huang2017adversarial}, C\&W \cite{carlini2017towards}, Deepfool \cite{moosavi2016deepfool}, PGD \cite{madry2017towards}, etc. The second one is \textit{black-box} attacks, where the adversary does not know any information about the target model. He can leverage the transferability property of AEs \cite{szegedy2013intriguing}, which refers to the ability of AEs generated from one model to attack another different model. The adversary can train a shadow model locally, and generate the corresponding AEs using conventional white-box attack techniques. Then these AEs have a high chance to succeed in attacking the target model. 

For black-box attacks, the attack success rate, i.e., AE transferability, highly depends on the similarity between the victim model and adversary's shadow model. Therefore, our model extraction technique provides a new opportunity of improving such similarity, thus the success rate of black-box attacks. In particular, the adversary can apply \Name to extract the architecture of the victim model, and then train a shadow model with this architecture. The AEs generated from this model enjoy higher transferability to the victim model compared to the ones from a random shadow model. 

\begin{figure}[t]

\subfloat[AE transferability\label{case1}]{
\centering
\includegraphics[width=0.45\linewidth]{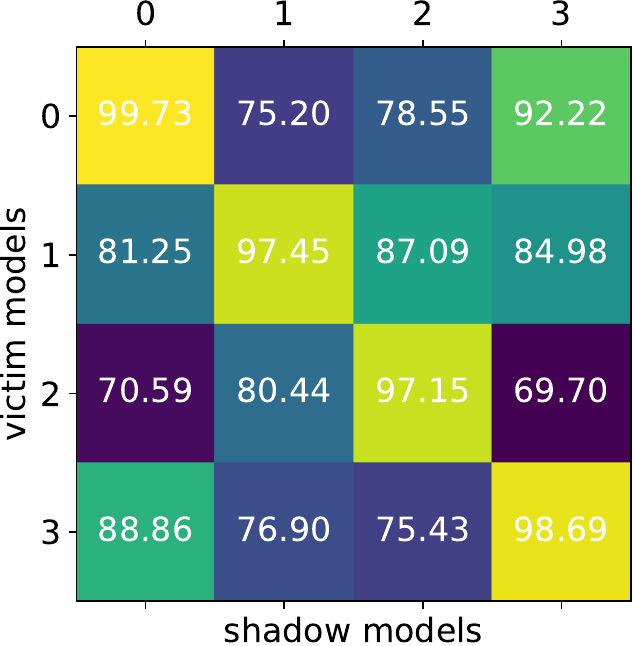}}
\centering
\subfloat[MIA\label{case2}]{
\includegraphics[width=0.45\linewidth]{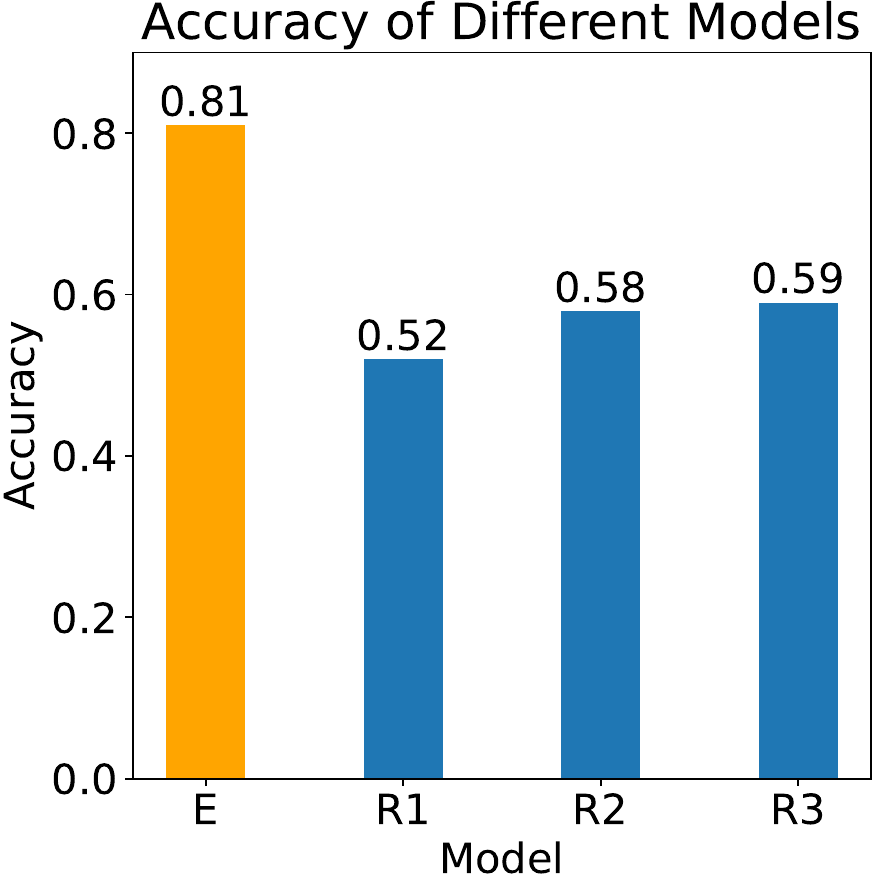}}

\caption{\Name enhances two attacks}
\vspace{-15pt}
\label{casestudy}
\end{figure}

Figure \ref{case1} validates the effectiveness of \Name in enhancing black-box adversarial attacks. The \textit{y}-axis represents four victim models with different architectures, while \textit{x}-axis represents the shadow models used for generating AEs with FGSM. The blocks with the same \textit{x} and \textit{y} indexes denote the success rates with the enhancement of \Name, while the rest blocks show the success rates with random architectures. For fair comparisons, we set the same perturbation scale (with $\epsilon=0.3$) for all cases. It is clear that AEs based on our \Name has the highest transferability (the diagonal blocks) for each victim model.

\subsection{Enhancing Membership Inference Attacks}
As the second case study, we show how \Name can facilitate the membership inference attack (MIA) \cite{shokri2017membership}. MIA is a type of privacy attack that aims to determine whether a particular data point has been used to train a machine learning model. This attack is of particular concerns in applications where the training data contain sensitive information, such as medical diagnoses, credit scoring, and fraud detection. To launch the MIA, the adversary can train multiple shadow models locally on separate datasets that are representatives of the target dataset. The adversary then queries these shadow models with a data sample and obtains the the corresponding predictions. These prediction vectors, along with whether the sample is a member of the training sets as the label, are used to train a membership inference model. With this model, the adversary can infer the membership of any sample based on the prediction results from the target model. 

Clearly the MIA accuracy is related to the similarity of the shadow models and target model: shadow models that are identical to the target can better reflect the attributes of the training data. In the black-box setting, our \Name can increase such similarity from the model architecture perspective. Figure \ref{case2} shows the MIA accuracy of four cases: shadow models with the architecture extracted by \Name (yellow bar, $E$), and with random architectures (blue bars, $R1 - R3$). All these attacks are carried out on the MNIST dataset. The target, shadow and attack inference models are trained on the datasets of sizes 2,500, 57,500 and 3,500, respectively. It is obvious that our extracted model yields the highest accuracy. 

\section{Related Works}
\label{sec:related}
\noindent\textbf{Model extraction attacks on DNN accelerators.}
As summarized in Table \ref{paper}, numerous attacks have been proposed to steal the model architecture or weights from the DNN accelerators. For instance,
Zhang et al. \cite{zhang2021stealing} used ROs to remotely steal the structure of an FPGA neural network. But it did not aim at real-world accelerator designs, and it only did layer tests instead of recovering the entire model.
Meyers et al.\cite{9786107} analyzed the impact of layer folding in accelerators, which makes attacks harder to perform. Subsequently, they showed how to recover the folding before the number of neurons. Gupta et al. \cite{gupta2023ai} launched the EM-based side-channel attack on NVDLA using  an oscilloscope. However, this attack requires physical access. Additionally, the attacker needs to manually split a full trace into trunks divided by different layers to train the model or initiate the attack through inference.
Tian et al. \cite{tian2021remote} proposed to remotely extract the architecture of the versatile tensor accelerator (VTA) using the TDC sensor. They manually distinguished different hyper-parameters by observing the distinct shapes in the power trace. The attack is realized in the layer-wise setting. Yoshida et al. \cite{yoshida2019model} introduced an attack to extract the model parameters from the EM leakage. Li et al. \cite{li2021power} initiated differential power analysis (DPA) attacks on 2D DNN accelerators to retrieve weights from the matrix multiplication accelerator. Yu et al. \cite{yu2020deepem} exploited EM side-channels to recover model architecture by simple EM analysis and then recover the model weights by adversarial learning. All the above three attacks require physical access to the victim device, and only target very simple accelerator implementations. 



\noindent\textbf{Other types of attacks on DNN accelerators.}
In addition to model extraction, model inversion attacks are also designed to recover the input inference samples. Wei et al. \cite{wei2018know} implemented a high-resolution oscilloscope to collect the power traces and recover the label of the input images. Moini et al. \cite{moini2021remote} designed a remote side-channel attack to recover MNIST images from a binarized neural network (BNN). Different from the above works, Luo et al. \cite{luo2021deepstrike} demonstrated an integrity threat to the DNN accelerators, where they performed a guided fault injection attack to alter the prediction of the victim model. Those attacks are outside the scope of this paper.


\section{Conclusion}
\label{sec:conclusion}
We propose \Name, the first automated remote side-channel attack against the Nvidia Deep Learning Accelerator (NVDLA). \Name leverages the readouts of TDC as the power indicator of NVDLA, and trains RNN-CTC and Transformer seq2seq models to predict the model architectures, and disclose the leakage points. Evaluation results demonstrate that \Name is able to extract the operation sequence of the victim model within an error rate of 1\%. It can also enhance existing black-box adversarial examples and membership inference attacks. 


\bibliography{mybib}{}
\bibliographystyle{IEEEtran}
\end{document}